\documentclass[authoryear,preprint2]{aastex}
\newcommand{\xC}{x_{\rm \scriptscriptstyle C}}
\newcommand{\yC}{y_{\rm \scriptscriptstyle C}}
\newcommand{\ldaC}{\lambda_{\rm \scriptscriptstyle C}}
\newcommand{\fC}{f_{\rm \scriptscriptstyle C}}
\newcommand{\xF}{x_{\rm \scriptscriptstyle F}}
\newcommand{\yF}{y_{\rm \scriptscriptstyle F}}
\newcommand{\RF}{R_{\rm \scriptscriptstyle F}}
\newcommand{\rhoF}{\rho_{\rm \scriptscriptstyle F}} 
\newcommand{\KD}{K_{\rm \scriptscriptstyle D}}
\newcommand{\Klam}{K_\lambda}
\newcommand{\bge}{\begin{equation}}
\newcommand{\ede}{\end{equation}}
\newcommand{\bga}{\begin{eqnarray}}
\newcommand{\eda}{\end{eqnarray}}
\newcommand{\wdh}{\widehat}
\begin{document}
\title{2D analytical modeling of distortion and sky background \\ in multi-fiber spectrographs: the case of the Norris spectrograph \\ at Palomar Mountain}
\author{M.\,Viton and B.\,Milliard}
\affil{\rm Laboratoire d'Astrophysique de Marseille, Traverse du Siphon B.P.8, 13376 Marseille cedex 12, France; maurice.viton@free.fr, bruno.milliard@oamp.fr \\ Received 2001 Oct. 17, accepted 2002 Oct. 30}
\begin{abstract}
	A method for optimal reduction of data taken with multi-fiber spectrographs is described, based on global correction of their geometrical distortion. Though it was specifically developed for reducing observations performed at Palomar Mountain using the Norris fiber spectrograph, this method can be adapted to other types of multi-object spectrographs such as the multi-slit ones. Combined with a 2D analytical interpolation of sky-background that accounts for non-uniform spectral resolution, the Norris software package achieves very high accuracy in airglow subtraction, even in the near infrared 
(7000-9000\,{\AA}) where molecular band-emissions commonly induce strong artefacts that preclude clean sky subtraction whenever standard image processing techniques are used. Correlatively, an improvement by a factor of 2 on the precision of radial velocities is achievable. Throughout the paper possible improvements to the method are suggested for those devising similar packages for other instruments. 
\end{abstract}

\section{ Introduction }

	In the last decade, multi-fiber spectrographs have opened a new era for performing efficient surveys of large and homogeneous samples of faint celestial sources. Combined with modern, large-format high quality CCDs, such systems allow (or will allow in the near future) simultaneous spectroscopy of sometimes as many as $\approx1000$ sources over fields of $\approx1$ square degree, that is $\approx$1-2 orders of magnitude more than is possible with multi-slit systems. However, two principal drawbacks of this powerful technology arise from:
\begin{enumerate}
\item The inevitably significant distortion of spectra over the CCD area, a property in common to all types of 
multi-object spectrographs. This distortion commonly results from : (i) the off-axis incidence of beams on the grating, thus inducing conical dispersion; (ii) the focusing optical components (collimator, imaging objective, etc). 
\item The poor performance of sky subtraction techniques in multi-fiber spectrographs, commonly resulting from: (i) the use of a limited number of sky fibers surrounding astronomical ones; and occasionally (ii) non-flatness of the CCDs. With little effect on distortion itself, this last cause can however be responsible for important variations in the bidimensional (2D) resolution, as shown in Sect.\,3. 
\end{enumerate} 

	Thorough analyses of various sky subtraction techniques applicable to fiber spectrographs and their capabilities to provide satisfactory signal-to-noise (S/N) ratios for very faint sources have been presented in several papers in the past decade, and useful references to them can be found in, for example, \citet{lis94} and \citet{wys92}, two of the most recent articles on the subject. However, as emphasized in particular by Wyse \& Gilmore, their precepts apply generally to $``$spectra already extracted from 2D raw data, after appropriate instrument specific pre-processing." Generally, the latter relies on use of a  standard method for correcting the marked distortion of large 2D-images, that is to say by fitting polynomials first to the transverse distortion of each spectrum, and then to the longitudinal (or dispersive) distortion with simultaneous conversion into wavelength units. And more critically, the coefficients of such fitted polynomials are allowed to be {\it independent}, i.e. not constrained to vary smoothly across adjacent spectra despite this being expected on physical grounds. So, when compounded with a resolution that normally varies across the field, this method usually leads to dramatic sky-subtraction residuals, especially in the red region of airglow emission overcrowded with intense OH lines. In addition, any instability in the geometrical distortion correction obviously results in uncontrolled systematic errors in the velocities derived from spectral lines. 

	In the course of spectroscopic follow-up of UV-excess sources detected at 2000\,{\AA} around the extragalactic cluster Abell 2111 by the balloon-borne FOCA experiment \citep{mil91},  we have been using the Norris multi-fiber spectrograph at Palomar Mountain Observatory. Because of the bright infrared sky, previous users of this instrument,  such as for example \citet[ hereafter SSH]{ssh97} did not attempt observations at wavelengths above 6500\,{\AA} where the adopted standard sky subtraction techniques proved quite unsatisfactory for fiber spectroscopy. Our first attempts at reducing our early data collected in 1995 this way confirmed that high-intensity residual $``$grass" effectively forbade identification of even moderately strong emission lines above 7000\,{\AA}, while the spectral domain required for our extragalactic program was $\simeq 3800-9000$\,{\AA}. 

	Investigation of the phenomenon soon suggested that it could result mainly from a combination of 
(i) insufficient accuracy in the wavelength calibration using independent fiber-to-fiber polynomial fits, as mentioned above, and (ii)  poor estimation of the precise sky spectrum to be subtracted from each fiber, because of the marked variations in resolution ($\simeq 3-5$ pixels) across the field, which essentially resulted from the non-flatness of the CCD used with the Norris spectrograph. Simulations show that both effects induce unacceptable artefacts in the sky subtraction (5 to 10\% of line peak intensity), when shifts in the position of lines or errors on the PSF width reach 0.05 or 0.1 pixel respectively.   

	So, we decided to spend time devising a special software package (based on specially written {\sc midas} procedures and {\sc fortran} subroutines) permitting an improved sky correction, thanks to (i) a 2D analytical model of the spectrograph's optical system that provides an accurate correction of the global distortion; and (ii) a 2D empirical model of the sky background contribution that ensures a reduced noise level by accounting for the non-uniform spectral resolution. This was non-trivial and hence the mathematical formulation of the algorithm has evolved markedly over the 3 years of data collection, leading to improved quality output spectra, which now show good S/N ratios even at their very red limit. 

	The multi-slit spectrographs are in principal free of such effects, at least in the ideal case where the slit length is much larger than the astronomical object. However, this condition is not always fulfilled during realistic observations, because maximizing the science return most often implies minimizing slit length to avoid overlapping spectral images. Hence, the accuracy of sky subtraction can be much reduced for such objects as well resolved galaxies or close galaxy pairs that are frequently revealed by spectroscopic follow-ups of deep imaging surveys. A global correction of the distortion and of the 2D sky contribution in multi-slits spectrograms may then improve the overall quality of the derived data, even using a model of the distortion simpler than described in this paper. 

	We know that some teams \citep[]{nor00}\footnote{http://obswww.unige.ch/Instruments/GIRAFFE/} are already working on  methods and software packages similar to the ones presented here. Although marked variations exist in the design of the multi-object spectrographs attached to the world's large aperture telescopes, we are confident that other people will be interested in adapting our method to their specific instruments, since in addition to spectra of much higher quality (which observers will fully appreciate), engineering teams can obtain better knowledge concerning the true limits of their spectrographs, as shown below. In this paper, we briefly report our observational technique at Palomar in Sect.\,2 and detail the modeling of the Norris system in Sect.\,3, while Sect.\,4 provides a short discussion of the performance and limits of our method, together with recommendations for further improvements. 

\section{ The Norris observations }

	The optical spectra of the UV-selected objects were obtained with the Norris multi-fiber spectrograph mounted at the Cassegrain f/16 focus of the Hale 5m telescope. The observations were performed  over 5 nights (out of a total allocation of  7) from 1995 to 1997. This  spectrograph has been extensively described by \citet{ham93} and was originally designed to position a maximum of 176 fibers (as close as $16''$) in a circular field of $20'$ diameter, using a 2048$\times$2048 CCD with 27$\mu$m pixels. However, our observations were carried out using the new Tektronix/SITe 2048$\times$2048 thinned, backside-illuminated CCD (hereafter  $``$CCD13") with 24$\mu$m pixels, which only enables $\approx$150 fibers to be imaged. This was no disadvantage for our programme since the number of UV-detected sources in a $20'$  circular field is normally $<50$. In fact, this has allowed us to improve the science return in several ways, in particular by assigning typically 20-35 sky fibers per field, from which accurate 2D fits of the night sky contribution to the observed spectra could be derived (see Sect.\,3 below). Note that the images taken with CCD13 consist of spectra with a mean FWHM of 4 pixels and spaced by 12 to 13 pixels.  
	
	The wavelength range for our 3 years of observations was $\simeq 5150$\,{\AA} wide, with edges near 3800\,{\AA} and 9000\,{\AA} but with minor shifts ($\simeq50-70$\,{\AA}) from year to year. This very wide spectral domain was obtained using a 300 line mm$^{-1}$ grating, with a lower wavelength limit requirement of observing not only the [O\,{\sc ii}] 3727\,{\AA} emission line of low-redshift field galaxies, but also the Ca\,{\sc ii} K (3933\,{\AA}) and H (3968\,{\AA}) absorption lines with virtually zero redshift, since $\approx 30\%$ of the UV-detections were expected to arise from galactic stars, as was confirmed by follow-up. With this grating, the dispersion was $\simeq$0.39\,pixel\,{\AA}$^{-1}$ but the spectral resolution varied from $\simeq$ 7 to 12\,{\AA} FWHM over the field, mostly due to a marked non-flatness of CCD13 (see Sect.\,3.2). Note also that the blaze angle being well optimized for these observations, no filter was needed to separate orders 1 and 2 over the spectral domain. 

	A nominal sequence of observation on a given $20'$ sub-field normally included two exposures of $50-60$\,minutes on the sky to facilitate cosmic ray removal, and each exposure was bracketed by 2 of the 3 short  exposures taken during the sequence with an FeAr arc for wavelength calibration. Flat-fielding was obtained through dome-flats taken with the telescope at the zenith, and using the $``$undithered" spectrograph mode (see SSH for details concerning the poor results obtained using the dithered mode, in which the collimator is moved back and forth  perpendicularly to the dispersion). However, CCD13, which we and SSH have used, is so uniform in sensitivity that little damage was expected a priori from minor shifts due to instrumental flexures in the position of spectra on the chip. Note that the effects of flexure on different parameters of the spectrograph are provided as a by-product of the method used for correcting the geometrical distortion 
(see below). For example, for the range of declinations that we have been using ($+29\degr \la \delta \la +43\degr$), the maximum shifts from the zenith (where dome-flats are taken) to hour angles as large as $60\degr$ remain $\la1$\,pixel (consistent with the Spectrograph Manual at Palomar), to be compared with the mean 4\,pixel FWHM of the Norris spectra. 

	Note that we have accounted for (i) the relative flux calibration using a standard star, (ii) the continuous telluric extinction at Palomar using the calibration of \citet{hay75}, and (iii) the molecular-band absorptions above 5500\,{\AA} using one of the program stars with low metallicity. More importantly for the present paper, though no velocity standard was observed,  58 galactic stars were identified around Abell 2111, the radial velocities of which have revealed (in addition to high-velocity halo stars) a population of disc and thick-disc objects with a mean velocity relative to the LSR close to zero, thus indicating that the derived velocities are of sufficient reliability for the needs of our programme. A further check was provided by 7 stars from another survey programme carried out at the same time around the globular cluster Messier\,92: 3 of them obviously belong to the cluster and show a mean difference of 
$-(10\pm26)$\,km\,s$^{-1}$ relative to the known heliocentric velocity of the cluster ($-121$\,km\,s$^{-1}$), a value not statistically different from zero (see Sect.\,4 for further information derived from spectra of galaxies). 

	It is here worth mentioning a positive aspect of fiber spectroscopy compared to classical slit or multi-slit spectroscopy,  where precise centering of sources is mandatory to avoid large errors in the derived velocities. Fibers normally eliminate such effects because the large number of internal reflexions lead to an angular output distribution of the beam highly $``$thermalized" relative to the input, that is, with no memory of any 
mis-centering of the input beam. 

	In both cases however, errors in relative flux calibration vs.\,wavelength remain a major problem when differential atmospheric refraction induces variable vignetting as a function of wavelength, and this applied to our follow-up  
since no atmospheric dispersion corrector was available at the Cassegrain focus of the Hale telescope. The correct positioning of the fibers therefore requires calculation of the refraction for the planned observation time. The representative wavelength at which this calculation is done must be chosen so as to equalize the chromatic offcentering at the two ends of the spectral domain. For the one we used, this optimal wavelength was $\simeq 4800-5000$\,{\AA}, but theoretically it is also a strong function of the ratio of the fiber width to the seeing. 

	Another observing constraint of critical importance at Palomar was to accurately check the centering of guide stars relative to several bright programme stars at the very beginning of each observing session, because of possible errors in the equatorial coordinates of the guide stars. Due to time constraints, this check was not performed for one of the subfields in 1996, which resulted in a  sensitivity loss of $\approx$1-2 magnitudes for some stars, but remained acceptable for extended objects such as galaxies (a main goal for our follow-up of UV-detected sources). 

\section{ Modeling the Norris system }

	In this section, we show how the geometry of the Norris system was used to derive the analytical formalism adopted for correction of the 2D spectrograms, and how the algorithm was applied. In addition to the homogeneous quality of results that one can expect from a global transformation of input data, this method also implies that images are resampled in only two steps (the 2D-correction itself, followed by wavelength rebinning of all 1D-spectra to a common scale, Sect.\,3.3), thus ensuring maximum conservation of the precious local S/N ratio 
and preserving spectral resolution. 

	With the aim of providing a general method for analytical modeling of various multi-object spectrographs, we give separate analyses for the different components of distortion (due to collimator, grating and camera) and the markedly variable 2D-resolution experienced on the Norris frames. While the former should adapt easily to other instruments whenever their optical layout is comparable to Fig.\,1 (upper diagram), the latter is more specific to the Norris spectrograph, and a different formalism may be required in other cases (for example: a reversed situation in which most of the variable resolution comes from the camera optics itself, instead from CCD non-flatness, as here). 
 
\subsection{ Basic geometrical formalism }

	For the reader's convenience, the upper diagram of Fig.\,1 partially replicates a meridian cut of the Norris optical train given by Hetal (their Fig.\,3), in which a flat grating is placed near the center of curvature G of a spherical collimator mirror, while the fiber bundle (or fiber slit) is placed at the collimator's focus F. The bundle is convex towards the spherical mirror as in any Schmidt design, and theoretically oriented perpendicular to the plane of the figure. At the focal ratio of beams exiting from  fibers ($\approx 4$), spherical aberration of the collimator is negligible. The parallel beams dispersed by the grating are then focused onto the CCD by a high-luminosity lens (Epps camera, see Hetal for details), the optical axis of which intercepts the CCD at point C.  
\begin{figure}[h]
\epsscale{1.0}
\plotone{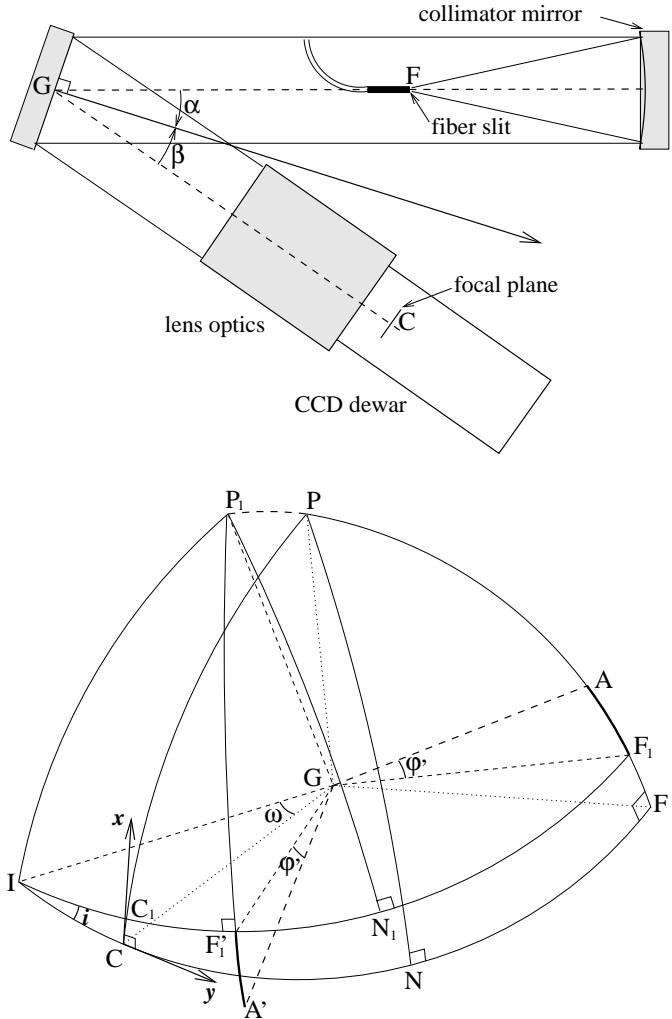}
\caption{ {\em Upper:} simplified meridian cut of the Norris optical train. {\em Lower:} sketch of the corresponding spherical triangles used for geometrical modeling. }
\end{figure}

	We must emphasize that {\em G has been implicitly assigned to the camera's optical axis} in our modeling of the system, which of course should be inexact in practice, and this might explain (at least partly) why (i) the position of C was found at pixel coordinates $\xC\simeq 1015$, $\yC\simeq 1000$ (instead of 1024, 1024 ideally); and (ii) some discrete, high-order and non-axisymmetric terms had to be introduced to correct small, but obvious residual trends in the observed distortion (at the level of $\ga 0.1$ pixel). However, another possible source of asymmetry will be mentioned below.  

	The lower diagram of Fig.\,1 provides a sketch of various spherical triangles involved in the derivation of trigonometrical formulae detailed below or in Appendix\,A. Ideally, the focal plane should be normal to $\vec{\rm GC}$ at point C  with the abscissae of spectra increasing along $\vec{\rm Cx}$ and wavelengths along $\vec{\rm Cy}$. Similarly, a parallel beam from a given fiber A is incident on the grating with field angle $\varphi=\wdh{\rm FGA}$, and the fiber bundle is theoretically perpendicular to $\vec{\rm GF}$ as mentioned above. 

	In the preliminary model devised for correcting the first series of spectra (from 1995), we had also implicitly assumed that the normal to grating, $\vec{\rm GN}$, belonged to meridian plane CGF, with angles $\alpha = \wdh{\rm NGF}$ and $\beta = \wdh{\rm NGC}$ accounting respectively for the angle of incidence of the collimated beam onto the grating and the angle of emergence for what is hereafter called the $``$central wavelength" $\ldaC$. Notice that so defined, $\alpha>0$ but $\beta<0$ in Fig.\,1. By mechanical construction, we have $\alpha-\beta = 35\degr$, but a value of $34\degr$ was erroneously adopted in 1995 and taken fixed in all versions of the software package. Analytical simulations show that the effect of this mistake is negligible on the wavelength distortion, but its possible contribution to the slight asymmetry of the chromatic distortion mentioned at the end of this section has not been checked.  

	The second series of spectra (from 1996) actually revealed a weak but continuous variation of the residual dispersion with $x$, leading us to suspect that the true normal to the grating, $\vec{\rm GN_1}$, did not lie in meridian plane CGF, but in a plane IGF$_1$ inclined at angle $i$ (taken $>0$ in Fig.\,1) to plane CGF. With both planes intersecting along a line $\vec{\rm GI}$ such that angle $\omega=\wdh{\rm IGC}$, cardinal points P,C,N,F of the theoretical optical system are transposed to  P$_1$,C$_1$,N$_1$,F$_1$ with point C$_1$ defined as lying in plane PGC, which is perpendicular to meridian plane CGF. Angles $\alpha$, $\beta$ now become $\alpha'$, $\beta'$ while $\varphi'=\wdh{\rm F_1GA}$ is the true input field angle of fiber A, and F$'_1$, A$'$ the focal plane images of F$_1$, A for wavelength $\lambda$. 

	By virtue of the conical dispersion law, we have also $\wdh{\rm F'_1GA'}=\varphi'$ while the true output dispersive angle is $\varepsilon'=\wdh{\rm C_1GF'_1}$ such that :
 \bge
 k n \lambda = \cos\varphi' \left[\sin\alpha' + \sin(\varepsilon'+\beta')\right]
\ede
with $n=300$ lines/mm and $k=1$. We omit  $k$ in what follows to simplify the formulae. Because our definition for central wavelength $\ldaC$ implies :
\bga
 \varphi' &=& \varepsilon' = 0, \cr
  n~\ldaC &=&  \sin\alpha' + \sin\beta', 
\eda
both dispersion formulae can be merged into a single one:	
\bga
 n \left(\lambda - \ldaC\cos\varphi'\right) + \sin\beta'\cos\varphi' =  \cr
 \sin\beta' \left[ \cos\varphi'\cos\varepsilon' \right] + \cos\beta' \left[ \cos\varphi'\sin\varepsilon'  \right] 
\eda
in which the two terms in brackets on the right-hand side depend on coordinates $xy$, the focal length $f$ of the camera, and angular parameters. However, the full derivation is somewhat complicated, so we detail it in Appendix\,A. It leads to the condensed parametric formulae:
\bga
{x\over f}&=&A_1\sin\varphi' \sqrt{1+{\rho^2\over f^2}} +A_2 {y\over f} + A_3, \cr
{y\over f}&=&B_1\sqrt{1+{\rho^2\over f^2}}\left[n\left(\lambda-\ldaC\cos\varphi'\right)+\sin\beta'\cos\varphi' \right] \cr
 &&~~~~~~~~~~~~~~~~~~~~~~~~~~~~~~+B_2 {x\over f} + B_3. 
\eda
Here $\rho=\sqrt{x^2+y^2}$ is the distance between points A$'$ and C, and $A_n$,$B_n$ are parameters which depend on angular variables only. Given that $i$ is normally expected to be very small, the exact formulae of Appendix\,A are well approximated by their first order terms:
\bga
 A_1 \simeq -1 &~~~~~& A_2 \simeq i~\cos\omega ~~~~~ A_3 \simeq i~\sin\omega,\cr
 B_1 \simeq {1\over \cos\beta'} &~~~~~& B_2 \simeq -i~\left(\sin\omega~\tan\beta'+\cos\omega\right), \cr
 &~~~~~& B_3 \simeq -\tan\beta' 
\eda
from which decisions on which way and by how much each of them has to be modified become obvious once a given set of parameters has revealed itself unsatisfactory and residual trends have been identified (see Appendix\,B for typical values of the parameters). 

	So far, we have analyzed the main components of geometrical distortion due to the collimator and the grating. However, early inspection of distortion-corrected FeAr calibration images revealed three minor additional effects :
\begin{itemize}
\item a slight, linear increase of ordinates $y$ (dispersive axis) vs. abscissae $x$ for fiber images for any given wavelength, thus implying a non-orthogonality of the fiber slit to meridian plane CGF, or a grating rotation in its plane, by an amount of $\approx15-20\,\arcmin$. 
\item a slight, linear variation of the apparent focal length $f$ (or dispersion) with abscissae $x$, also implying a tilt of the fiber slit in plane PGF of $\approx 13\,\arcmin$.
\end{itemize}
Both effects have been simply accounted for by parameters with dimensions of slopes (respectively {\sc slop} and {\sc tilt}, see Appendix\,B). Interestingly, it transpired that their values varied slightly with telescope position on the sky, in a way consistent with differential flexure of some component inside Norris. 
\begin{itemize} 
\item a marked trend of residual distortion vs.\,radial distance from point C, which is well approximated by a degree-4 polynomial in $\rho$. An initial assumption was to suspect a dominant contribution from the camera's own distortion, but this proved  inconsistent with later information that its distortion law is essentially of degree 2 (Hetal). In addition, further analysis led us  to suspect that the degree-4 term resulted from CCD13's non-flatness, which has little effect on distortion itself, but deep implications for the sky-background subtraction. 
\end{itemize}

	Modeling of the camera's distortion proved the most critical point of the method, and certainly the most subject to marked evolution in its mathematical formalism over the 3 years of data reductions. The formulae presented here are the final solutions adopted on the basis of critical information reported by Hetal in particular. They consist first in a coefficient for radial distortion $\KD$ and second in a chromatic term $\Klam$. $\KD$ was written as a polynomial function of the field angle of beams exiting from the camera lens expressed through its tangent $\rho/f$: 
\bga
 \KD = 1+\sum_{n=1}^4 O_n (\rho/f)^n
\eda
since theoretically this is a better representation than using a dependency on $\rho$ alone, given that focal length $f$ can vary slightly over time for various reasons (instrumental setup, flexure and/or thermal gradients). However, for our observations with the Norris spectrograph, $f$ was found to vary by only $\approx 0.13\%$ peak-to-peak and hence this formulation might be seen as unnecessarily finicky. The values derived for the $O_n$ coefficients confirm a dominant contribution ($>97\%$) from the degree-2 term, and the overall contribution of the camera accounts for $\Delta\rho\simeq+14$ pixels at the extreme corners of the field, i.e. most of the observed distortion. 

	It was found necessary to include the chromatic term $\Klam$ as an additive term to $\KD$, this is ascribed to the intrinsic chromatism of the camera. It was arbitrarily accounted for by a degree-5 polynomial with numerical coefficients that were derived from fits to residual trends. Note that the maximum contribution observed for $\Klam$ was $\approx 0.6$ pixel in $x$ and $\approx 0.4$ pixel in $y$.  Though such values are roughly consistent with data reported by Hetal,  a possible cause  for this asymmetry has been mentioned above.  

\subsection{ Variable resolution and true sky-background }

	Numerical data on the 2D-profile of CCD13's $``$bump" kindly provided by T.A.\,Small (1997, private communication) prompted us to think that the bump might be simultaneously responsible for (i) most of the marked variations observed in the $x$ and $y$ linear resolutions; and (ii) most of the degree-4 trend in $\rho$ mentioned above. Analytical modeling confirmed that the bump is not symmetrical, but can be acceptably approximated by elliptical contours oblate in $y$, with an axial ratio $\simeq 0.87$ and a center at pixel coordinates $\xF\simeq 1000\pm 2$, $\yF\simeq 1196\pm 10$ (standard errors on the mean), a fair distance away from point C. Combined with the slightly curved focal plane of the camera ($\le 50\mu$m) but in the opposite sense according to Hetal (the CCD bump being convex towards camera lens), the global amplitude of $\approx (320\pm 20)\,\mu$m (i.e. 12-14 pixels) derived from our modeling is quite consistent with a contribution from CCD13 of $\approx 280\,\mu$m measured by T.A.\,Small. 

	Focusing of the Norris spectrograph itself is achieved by moving the collimator, and focus adjustment is  performed at the beginning of observing sessions. The best compromise consists of course in roughly equalizing the intra- and extra-focal amounts of defocus, which normally correspond to a locus of best focus with a radius of $\approx 800-900$ pixels, where the FWHM of the point spread function (PSF) is at a minimum of $\simeq 3.1$ pixels; while the FWHM typically reaches $\approx 5$ pixels at the bump center or at the CCD corners.  

	Tests showed that in addition to a nearly constant term, an even polynomial of degree 4 with constrained coefficients was sufficient to account for the varying FWHM $w(x,y)$ due to this effect, that is:
\bga 
 w(x,y) = \gamma \times \rhoF^2 \left( 1-{\rhoF^2 \over 2\RF^2} \right) + c(x,y)
\eda
in which $\RF$ refers to the radius of an ideal circle, centered at pixels coordinates $\xF,\yF$, around which focusing would be  best. As mentioned above, this locus being neatly elliptical, $\rhoF$ has been formulated so as to include a factor $h_y$ ($>1$ hereafter) for homothetic transformation along the $y$-axis, such as:
\bga 
 \rhoF^2  = \left(x-\xF\right)^2 + h_y^2\times \left(y-\yF\right)^2.
\eda
The $``$nearly constant term" $c(x,y)$ has actually been fitted by a plane, i.e. $c_1+c_2 x+c_3 y$, with very small values observed for $c_2$ and $c_3$, corresponding to non-orthogonality of the mean CCD plane with respect to the camera optical axis by $\approx 15\arcsec$ (a quite satisfactory figure, though probably meaningless in terms of true optical misalignement given the irregular shape of CCD13).  

	Notice that this formulation for $w(x,y)$ implies (i) that $\gamma<0$ and hence $c(\xF,\yF)$ is accounting for the width of the PSF at the bump's center; and (ii) that the global shape of the curved CCD13 surface is close to an oblate paraboloid,  
which is far from being realistic according to the data on its true shape. So, the formulation adopted here leads to a model PSF width which is systematically in error in some areas of the field (see Sect.\,4). 

	 As emphasized in Sect.\,1, marked variations in the 2D-resolution of multi-object spectrographs can be responsible for dramatic losses of accuracy in the subtraction of sky backgrounds, and this whatever the method used for optimal subtraction, if interpolation of the background does not account for local variations in the PSF. In addition, it can be shown that whenever the monochromatic PSF remains approximately Gaussian (which applies to Norris spectra, with nearly the same FWHM values in $x$ and $y$), slight loss of focus induces variations in local intensities which are also well accounted for by an even polynomial of degree 4 with the same formalism as the first term in Eq.\,7. 

	The four shape parameters ($\xF, \yF, \RF, h_y$) were used in a {\sc fortran} subroutine ({\sc spfndfib}, see Fig.\,2) for interpolating sky-fibers' intensities row by row. However, given the limited number of sky fibers and the potentially low S/N ratio of faint signals, spurious intensity values were eliminated iteratively by automatic weighting for each row. 

\subsection{ Method of reduction and software application }

	As mentioned in Sect.\,1, we have been using the {\sc midas} package (more exactly the old 1994 version for VMS 
$\alpha$-stations) as a basic tool for devising the series of specific procedures required for complete reduction of the Norris images. Figure\,2 provides a flow chart of the essential ones and how they proceed automatically or with interactive control by user, but more details on some critical points of our method are provided here. 
 
	Primary extraction of data was achieved by {\sc extract1}, with in particular two important steps to notice: 
\begin{description}
\item a) Cosmic rays 
\end{description}
High intensity hits (mostly due to electrons) have been removed by local detection (rather than from a comparison of 2 exposures) over boxes of $3\times3$ pixels using the {\sc rcosmic} command, based on a median filter with relative threshold and no inclusion of the central pixel. Care has been taken not to overcorrect for cosmic ray hits in choosing the threshold level, which would result in negative artefacts and hence a loss of useful signal, especially when hits occur on a steep gradient of intensity in the spectrum. So, weak intensity hits (mostly due to muons) frequently remain unaffected by {\sc extract1}, but they can be removed manually in the last stage of reductions using {\sc cumul}
\begin{description}
\item b) Stray-light background
\end{description}
This was fitted in the traditional way as the lower enveloppe to the intensity distribution of fibers in the astronomical and mean 
dome-flat images. A median filter of 3 pixels in $x$ by 33 pixels in $y$ was first applied and then minimum intensities found in boxes of 64$\times$64 pixels were fitted with a 2D cubic spline. Two iterations were performed to improve the overall accuracy of background determination. Given that the mean spacing between fibers is $12-13$ pixels on Norris frames, this technique has proven fairly efficient in general, except when  some fibers were assigned to very bright programme stars ($B\approx 13-14$), or more significantly for exposures taken close to the bright globular cluster Messier\,92 (see Sect.\,2). 

	The method used for correcting the geometrical distortion consisted basically of four steps: 
\onecolumn
\begin{figure}[h]
\epsscale{0.99}
\plotone{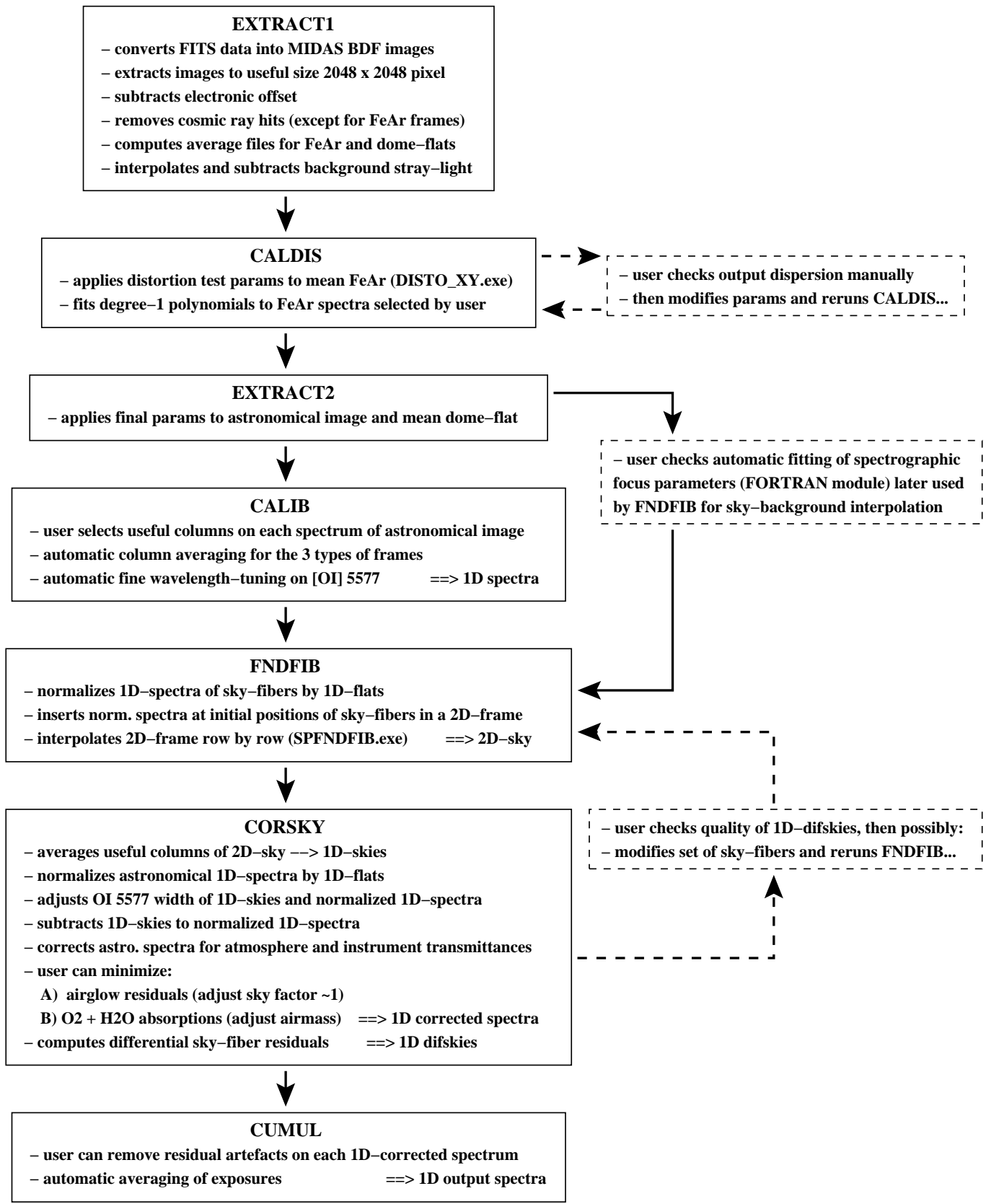}
\caption{ Flow chart of the software package. Specific {\sc fortran} subroutines (compiled in {\sc midas}) called by some procedures are indicated. Dashed arrows refer to tests (frequently iterative) performed by user. }
\end{figure}
\twocolumn
\begin{enumerate}
\item start from a blank output image (OI) 
\item apply the geometric distortion algorithm to OI pixel coordinates $xy$ using crude estimates of initial parameters, and get distorted pixel positions $x'y'$ 
\item fill the OI with intensities interpolated at positions $x'y'$ of the mean FeAr calibration image 
\item test for residual distortions in the OI, and iterate on parameters until satisfied.  
\end{enumerate}
Steps 1 to 3 are performed by the {\sc caldis} procedure with calls to the {\sc disto$\_$xy} procedure, a  modified version of the standard {\sc rebin/rotate} command for image rebinning, the core {\sc fortran} subroutine of which ({\sc rotait}) is  implemented with formulae given above and the set of model parameters (see Appendix\,B) stored in a special keyword for calls to {\sc disto$\_$xy}. The user is then prompted: (i) to choose a central fiber with good S/N ratio; (ii) to identify 2 reference lines (Ar{\sc ii} 4132\,{\AA} and Ar{\sc i} 8521\,{\AA}) for temporary linear wavelength rebinning; (iii) to select a number of fibers (typically 50) with good S/N ratios and well distributed over the whole field. {\sc caldis} then takes averages of useful columns for each selected fiber, adjusts Gaussian profiles to all calibration lines (typically 25), fits degree-1 polynomials to the dispersion, and plots calculated wavelength differences vs.\,calibration wavelength, so that the user can eliminate any unreliable lines. Finally, the distribution of mean dispersion is plotted vs.\,fiber position, and all data useful for the user's analysis of the residual distortion in step 4 are stored in a series of tables and descriptors.  

	Step 4 usually was the most difficult part of the data reductions and consisted in a thorough examination of the corrected FeAr image, with a variety of checks using either standard {\sc midas} commands or various procedures specially devised for determining residual trends, followed by decisions as to  which way and by how much the parameters were to be modified. (We do not dare to detail here the whole set of tests applied, since they depend strongly on the underlying package used for image processing). 

	Another critical step in data reductions consisted in carefully checking the quality of the 2D-sky interpolated by analytical modeling of the local PSF, as described above. The process was iterative in most cases and based on visual inspection of the sky-subtracted 1D-spectra obtained for all fibers after application of the {\sc corsky} procedure (Fig.\,2). As is commonly observed in deep spectroscopic surveys, some of the assigned sky fibers can then appear polluted by weak sources and must be removed from the set of fibers used for deriving the 2D-sky (serendipitous evidence of unexpected sources of interest are not uncommon in this respect). Conversely, some of the fibers ascribed to programme sources actually can show zero signal once sky-subtracted and hence should be used as true sky fibers in the next iteration (two iterations were generally sufficient for our data). 

	The very final step consisted of averaging the two exposures normally acquired successively on a given field (this is 
undoubtedly the stage of reductions prefered by most observers!) using the {\sc cumul} procedure, possibly after the user has replaced spurious artefacts on each spectrum by linear interpolation over as few pixels as possible (such artefacts may be due to faint cosmic rays incompletely removed by {\sc extract}1, or more frequently to strong airglow lines such as [O{\sc i}] 5577\,{\AA} and 6300\,{\AA}). 

\section{  Performance summary }

	A number of critical points have already been mentioned in sections above, with recommendations sometimes given for improvements to the similar methods now in use or development elsewhere. So, in this last section let us concentrate on more specific aspects together with quantitative indications of the performances attained by us in correcting the geometrical distortion and  modeling the variations of local PSF.  
\begin{description}
\item a) Geometrical distortion 
\end{description}
Uncorrected residuals in the distortion can be quantified in two ways: (i) transverse residual distortions vs. fiber position $x$ on the CCD were generally measured as $<0.2$ pixel during final examination of FeAr images corrected by {\sc caldis} (step\,4 of Sect.\,3.3), consistent with relative variations of $\la 5\times10^{-5}$ observed in plots that were considered satisfactory of the mean dispersion vs.\,$x$ ; (ii) velocities derived from well-separated lines over the spectral domain for stars (absorption lines) and galaxies (narrow emission lines) generally show variations of $<30\,{\rm km/s}$, quite suitable to the needs of our extragalactic program. 

	It has been shown in Sect.\,2 that the velocities derived for a series of galactic stars and also for 3 stars around the globular cluster M92 indicated a low systematic error, consistent with reasonable correction of the distortion, and with rms errors of $\simeq25-30\,{\rm km/s}$. This value is similar to the rms internal error per line for our spectra with high S/N ratio, and also consistent with the $\simeq35\,{\rm km/s}$ rms dispersion observed in the differential velocities obtained for a series of 8 galaxies for which 2 or 3 different spectra are available (due to overlaps between different subfields observed over the 3 years of follow-up). Based on a comparison with observations performed by the referee (Donald Hamilton, private communication) using the Norris spectrograph with a 1200 line/mm grating (instead of 300 here), our $25-30\,{\rm km/s}$ rms error is also consistent with a factor of 2 improvement on the accuracy of velocities, assuming the same S/N ratio of data.  Such accuracy is of great importance for studies now in progress concerning the nature of the stellar objects with UV-excess detected by FOCA, and of course it results directly from our method for correcting the global distortion of the spectrograph. 
  
\begin{description}
\item b) PSF modeling 
\end{description}
Ideally, a major improvement of our method concerning the model PSF would have consisted in using a 
2D-polynomial fit to CCD13's true shape (see Sect.\,3.2), but for various reasons this was not possible when the reductions were performed. Figure\,3 is a contour plot recently obtained for the calculated difference between: (i) the width of the PSF that would result from the bump shape itself; and (ii) the width resulting from an oblate paraboloid similar to typical fits  derived for $w(x,y)$ (though of course excluding the proper contribution of the camera lens to the focal plane curvature), i.e. using a locus of best focus defined by $\RF=900$ pixels and $h_y=1.13$ (dashed ellipse). Variations in the width of the PSF vs. defocus have been calculated using an effective focal ratio of 2.7 (according to numerical simulations) for Gaussian monochromatic beams exiting from the camera optics, instead of $\simeq2.0$ optically. 
\begin{figure}[h]
\epsscale{1.0}
\plotone{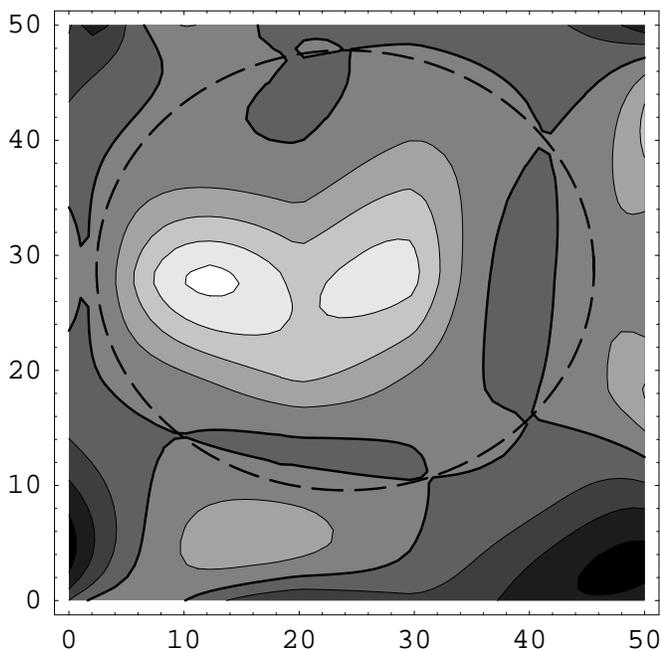}
\caption{ Contour plot of the difference between the true width of the local PSF due to the bump in CCD13 and an oblate paraboloid similar to typical ones derived from analytical modeling (Sect.\,3.2). Contour levels are in pixel units and range from $-0.45$ (dark) to $+0.60$ (light) in steps of 0.15 pixel ({\em zero contour} level is fatter). CCD $xy$ coordinates are in mm. The {\em dashed ellipse} shows a typical locus of best focus with $\RF=900$ pixels and $h_y=1.13$, along which the PSF is $\simeq 3$ pixels FWHM. The $x$ coordinate of fibers used for recording the spectra of galaxies shown in Fig.\,4 are indicated by arrow heads, with galaxies labels at the top of figure (red limit of spectral domain), and related comments in Sect.\,4b. }
\end{figure}

	The contour levels show that the width of the model PSF may have been in error by as much as $+0.5$ pixels (lower right corner) and $-0.6$  pixels (center left), i.e. $\lesssim20\%$, but in very small areas of the field. However, that $\RF$ and $h_y$ are physically related was not understood when reductions were made, and slightly erroneous values of $h_y$ were  sometimes used. This has certainly contributed to the moderate success observed in the subtraction of the infrared 
sky-background for some spectra, even with the 1997 version of the package and reduction method. So, we strongly recommend the use of more realistic analytical approximations to $w(x,y)$ for modeling other instruments whenever possible. 
\begin{figure}[h]
\epsscale{1.0}
\plotone{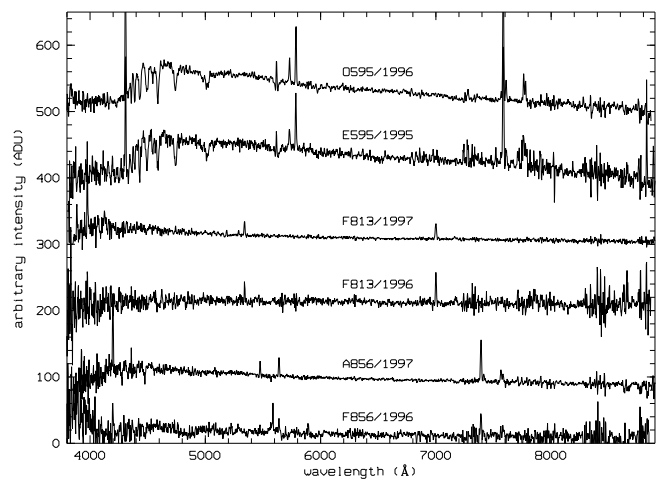}
\caption{ Examples of progress from year to year in the accuracy of geometrical correction and sky-subtraction for 3 galaxies, each observed twice with the Norris spectrograph. Letters (A, E, F, O) in front of FOCA's ID numbers denote subfields, and years of observation are indicated. Intensities have been offset for presentation, but true values of continua around 6500\,{\AA} are: $\approx15$\,ADU for IDs. 813 and 856, and $\approx85$ and 150\,ADU respectively for spectra E595 and O595 (this latter has been rescaled to the mean intensity of E595 before offsetting, so as to show spectral features with comparable intensities). A minimum sky contribution of $\approx 25$ ADU was typically subtracted around 6500\,{\AA}. See Sect.\,4 for comments on residual artefacts in the infrared or at the UV-cutoff of the spectrograph's spectral sensitivity. }
\end{figure}

	In any case, the unoptimized method we have used has given satisfactory results as is shown by Fig.\,4, a plot of 6 spectra obtained for 3 of the 8 galaxies mentioned above with low to moderate continuum intensities. These selected spectra provide examples of the progress experienced over the years in our understanding of the Norris system, and show that the sky-subtraction accuracy is essentially dependent on it, and more discretely on the accuracy of PSF modeling: 
\begin{enumerate}
\item the 2 spectra obtained one year apart for \#813 were recorded at the same $x$ position (Fig.\,3), a region where residuals between the true shape of CCD13 and the model fit are near minimum. So, the much better spectrum obtained in 1997 (especially above 7000\,{\AA}) is consistent with a globally better fit to the distortion, since with a radius of best focus  $\RF\simeq 1086$ pixels in 1997 (instead of $\approx 745$ in 1996), a much better fit to the local PSF is expected at the blue and red limits as mentioned in Sect.\,3.2. Note in conjunction the very low intensities of the observed continua ($\approx 15$ ADU) for  \#813 and 856, to be compared with the minimum sky contribution (i.e. the pseudo-continuum at a resolution of about 10\,{\AA}) of $\approx 25$ ADU typical around 6500\,{\AA} 
\item \#856 is also a weak continuum galaxy, but observed with 2 different $x$ positions one year apart. The later A856 spectrum is far better than F856 of 1996, but the sky subtraction was less accurate for A856 than for F813 in 1997 (though $\RF$ was nearly the same in both cases), because the $h_y$  factor was erroneously fixed at 1.14 for A856, as mentioned earlier
\item \#595 is a galaxy with recorded intensities much higher than for the previous cases. The O595 spectrum taken in 1996 benifited from much better modeling of geometrical distortion, and though it was recorded at a position where gradients in the PSF are a maximum, it shows a low level of residual $``$grass" above 7800\,{\AA}, while the E595 spectrum taken in 1995 appears badly corrected above 7200\,{\AA} despite its optimal position on the CCD. 
\end{enumerate}

	Another way to improve the accuracy of our sky-background subtraction would have consisted in positioning the sky fibers as uniformly as possible in the {\em output} focal plane, so as the row-by-row interpolation described in Sect.\,3.2  would be optimally constrained. However, this was not really the case for our observations, since the sky fibers were simply placed in areas of telescope's focal plane void of programme sources, partly because of mechanical limitations proper to the Norris system, but mostly because optimization of their output distribution would have required a great effort planning the setups. 

	Concerning the low S/N ratios and obvious systematic trends obtained at the blue limit of our spectral domain, they result essentially from a combination of: (i) a marked drop in the overall transmittance of the system for wavelengths shorter than 4000\,{\AA}; (ii) an even more marked drop in the spectral flux intensity of dome-flats because of their intrinsically very red colour; and (iii) the inadequacy of the instrumental spectral response derived from spectra taken of the calibration star, mostly because of seeing-dependent factors. 

     Let us also mention that a marked fringing above 8000\,{\AA} has been sometimes reported by observers using 
CCD13, but no such effect was conspicuous in our Norris images, though good spectra with high S/N ratio at such  wavelengths have been obtained for a series of bright programme stars. 

	Finally, we have found  the Norris spectrograph a powerfull instrument, though it is not often used. With the method of reduction described in this paper, the overall percentage of success rate for identifying the nature of the UV-sources detected by FOCA around Abell 2111 has been as high as $95\%$ globally (against $60\%$ using standard techniques as tested in 1995), an excellent performance given the dominant contribution from galaxies with blue magnitudes in the range 18 to 21. 
\acknowledgments  

We are greatly indebted to: Christopher Martin (at Caltech) for the nice opportunity of observing with the $``$Big Eye" at Palomar Mountain; and Todd A. Small (now at Caltech) for his efficient assistance during the three observing runs, critical data on the shape of CCD13 and general information on the Norris spectrograph. We also wish to thank: The Caltech and Palomar engineers for their support and most especially Eric Bloemhof; Antoine Llebaria at LAM for his creation of the {\sc fortran} sky-background interpolating subroutine {\sc spfndfib}, which had a decisive impact on the overall success of our package; V\'eronique Cayatte and Pierre North at ESO for fruitful  discussions during preparation of the {\sc giraffe} software package; WilliamTobin (University of Canterbury, NZ) for revision of the English of this paper; and Donald Hamilton (the referee) for a pertinent estimate of the improvement in velocity resolution achieved by our method, in addition to his thorough analysis of the manuscript.

\appendix
\section{ Derivation of the $A_n B_n$ parameters }
	Adopt: $\theta = \wdh{\rm C_1CA'}$ the position angle of point A$'$ a distance $\rho$ from center C  in the ideal focal plane; $\chi=\wdh{\rm CGC_1}$ the angular distance of C$_1$ from C as seen from grating's center; and angle $\eta=\wdh{\rm CC_1I}$, such that: 
\bga
 \tan\chi  &=& \tan i\sin\omega, \cr
 \cos\eta &=& \sin i\cos\omega, \cr
 \sin\eta  &=& \cos i\cos\chi.
\eda
The following relationships apply to spherical triangles C$_1$CA$'$ and C$_1$F$'_1$A$'$: 
\bga
 x &=& \rho\cos\theta, \cr
 y &=& \rho\sin\theta, 
\eda
\bga
 \sin(\wdh{\rm CGA'})  &=& {{\rho/f}\over \sqrt{1+{\rho^2/f^2}}}, \cr
 \cos(\wdh{\rm CGA'}) &=& {1\over \sqrt{1+{\rho^2/f^2}}}, 
\eda
\bga
 \sin(\wdh{\rm C_1GA'}) &=& \sin(\wdh{\rm CGA'})\sin\theta / \sin(\wdh{\rm CC_1A'}) \cr
                                              &=& \sin\varphi'/\sin(\eta+\wdh{\rm CC_1A'}), 
\eda
\bga
 \sin\theta\cot(\wdh{\rm CC_1A'}) = \sin\chi\cot(\wdh{\rm CGA'}) - \cos\chi\cos\theta, \hbox{~~~~} \cr
 &&
\eda
\bga
 \sin\varphi' &=& \sin(\wdh{\rm CGA'})\left[\sin\eta\sin\theta\cot(\wdh{\rm CC_1A'})+\cos\eta\sin\theta\right] \cr
                      &=& { {\sin\eta\sin\chi - (x/f)\sin\eta\cos\chi + (y/f)\cos\eta} \over \sqrt{1+{\rho^2/f^2}}}. \cr 
                       &&
\eda
The two terms in brackets of Eq.\,3 can then be expressed as functions of useful variables only: 
\bga
 \cos\varphi'\cos\varepsilon' &=& \cos(\wdh{\rm C_1GA'}) \cr
      &=& \cos\chi\cos(\wdh{\rm CGA'}) + \sin\chi\sin(\wdh{\rm CGA'})\cos\theta \cr
      &=& {{\cos\chi + (x/f)\sin\chi}\over \sqrt{1+{\rho^2/f^2}}}, 
\eda
\bga
 \cos\varphi'\sin\varepsilon' &=& \sin(\wdh{\rm C_1GA'}) \cos(\pi-\eta-\wdh{\rm CC_1A'}) \cr
      &=& \left[\sin\eta - {\cos\eta \over \tan(\wdh{\rm CC_1A'})}\right] {y/f \over \sqrt{1+{\rho^2/f^2}}} \cr
      &=& {{(y/f)\sin\eta + (x/f)\cos\chi\cos\eta - \sin\chi\cos\eta} \over \sqrt{1+{\rho^2/f^2}}}, \cr
       &&
\eda
from which the $A_n,B_n$ parameters of Eq.\,4 are derived:
\bga
 A_1 &=& -{1\over \cos\chi\sin\eta}, \hbox{~~~} A_2 = +{1\over \cos\chi\tan\eta}, \hbox{~~~}  A_3 = +\tan\chi, \cr
 B_1 &=& +{1\over \cos\beta'\sin\eta}, \hbox{~~~}  B_2 = -\left({\sin\chi\over \sin\eta}\tan\beta' + {\cos\chi\over \tan\eta}\right), \cr
 B_3 &=& -\left( {\cos\chi\over \sin\eta}\tan\beta' - {\sin\chi\over \tan\eta} \right). 
\eda

\section{ Set of model  parameters for geometrical correction }

	We detail here the nature of the adjustable parameters only (i.e. excluding the chromatic term $\Klam$, see Sect.\,3.1), and comment briefly on their values and usefulness in some cases. The 15 parameters used in the 1997 version of the package were classed into four groups corresponding to:
\begin{description}
\item 1) the fiber-slit
\end{description}
{\sc slop} \phantom{xxxx} tangent of the angle between the slit and plane PGF, hence $y'=y+${\sc slop}$\times (x-\xC)$ \\
{\sc tilt} \phantom{ixxxx} tangent of inclination in this plane
\begin{description}
\item 2) the grating
\end{description}
{\sc incl} \phantom{ixxxx} angle $i$. Found $\simeq-0\fdg2$ in 1996, but fixed to zero in 1997. \\
{\sc omeg} \phantom{ixxx} angle $\omega$. Found $\simeq-19\degr$ in 1996, but fixed to zero in 1997. \\
{\sc beta} \phantom{xxxx} angle $\beta'$  \\
{\sc ldac} \phantom{xxxx} central wavelength $\ldaC$ \\
The anomalous values observed for {\sc incl} and {\sc omeg} in 1996 probably resulted from accidental 
mis-positioning of grating's mount during adjustment of the spectral domain. 
\begin{description}
\item 3) the camera lens
\end{description}
$\xC, \yC$ \phantom{xxx} coordinates of optical axis \\
{\sc fobj} \phantom{xxxx} paraxial focal length $\fC$, hence true focal length $f=\fC+${\sc tilt}$\times (x-\xC)$ \\
$O_n$ \phantom{ixxxxx} distortion coefficients ($n\le4$) \\
The values derived for {\sc fobj} revealed sensitive to temperature in the dome, and  roughly consistent with thermal expansion factors of common metals. 
\begin{description}
\item 4) the CCD
\end{description}
{\sc rota} \phantom{xxxx} $y$-axis orientation relative to meridian plane CGF. Year dependent. \\
{\sc pent} \phantom{xxxx} accounts for inclination in this plane, hence $y'=y\times(1+${\sc pent}). Found irrelevant $>1995$,  fixed to null. 

	So, though only 12 active parameters were used for geometrical modeling of the Norris system in 1997, the formalism developed in Sect.\,3.1 and Appendix\,A allows treatment of more complex situations for other instruments in which parameters such as {\sc incl} and {\sc omeg} differ significantly from zero. 

\end{document}